\documentstyle[seceq,epsf]{ptptex}

\title{
Numerical Simulation of Interaction between an L1 Stream and
an Accretion Disk in a Close Binary System. 
}

\author{
Hidekazu {\sc Fujiwara}, Makoto {\sc Makita}, Takizo {\sc Nagae}
and Takuya {\sc Matsuda}
}

\inst{
Department of Earth and Planetary Sciences,
Kobe University, Kobe 657-8501}

\recdate{
}

\abst{
The hydrodynamic behavior of an accretion disk in a close binary
system is numerically simulated.
Calculation is made for a region including the compact star and the
gas-supplying companion. The equation of state is that of an ideal
gas characterized by the specific heat ratio $\gamma$.
Two cases with $\gamma$ of 1.01 and 1.2 are
studied. Our calculations show that the gas, flowing from the
companion via a Lagrangian L1 point towards the accretion disk, forms
a fine gas beam (L1 stream), which penetrates into the disk. No hot
spot therefore forms in these calculations. Another fact discovered
is that the gas rotating with the disk forms, on collision with the L1
stream, a bow shock wave, which may be called an L1 shock. 
The disk becomes hot because the L1 shock heats the disk gas in the
outer parts of the disk, so that the spiral shocks wind loosely
even with $\gamma=1.01$. The L1 shock enhances the non-axisymmetry
of the density distribution in the disk, and therefore the angular
momentum transfer by the tidal torque works more effectively.
The maximum of the effective $\alpha$ becomes $\sim 0.3$.
The 'hot spot' is not formed in our simulations, but our results
suggest the formation of the 'hot line', which is the L1 shock
elongated along the penetrating L1 stream.
}

\begin{document}

\maketitle

\section{Introduction and Summary}
\subsection{Discovery of Spiral Structure}
In recent years, Steeghs, Harlaftis \& Horne\cite{rf:1} discovered, with
use of the Doppler tomography technique, a spiral structure in the
accretion disk of the dwarf nova IP Pegasi, occurring in its outburst
phase. After the discovery of Steeghs et al.,\cite{rf:1} \ similar 
spiral structures have been found on many other accretion disks.
\cite{rf:2,rf:3,rf:4} \ Such spiral structures are generally
thought to be unobservable during quiescent phases of dwarf novae,
but Neustroev \& Borisov\cite{rf:5} observed them in a quiescent
phase in U Gem.

Formation of spiral-shape shocks in an accretion disk of a close
binary had been predicted based on two-dimensional numerical
simulations, by Sawada, Matsuda \& Hachisu,\cite{rf:6,rf:7} \ and 
Sawada et al.\cite{rf:8} \  Similar two-dimensional numerical 
simulations performed
by many researchers thereafter all confirmed the presence of the
spiral shock waves.\cite{rf:9}\tocite{rf:14} \ Self-similar solutions 
of the spiral shocks were also obtained.\cite{rf:15} \ Angular 
momentum loss at the spiral shocks provides a
different mechanism from the conventional $\alpha$-viscosity one. 

Three-dimensional simulations have also been made on the same
object, but given no results completely agreeing with one another.
Most of the three-dimensional numerical simulations of accretion disks
so far performed used the particle methods such as the SPH 
method,\cite{rf:16}\tocite{rf:23} \ and mostly failed in finding any
spiral shock.

On the other hand, Yukawa, Matsuda \& Boffin\cite{rf:24} observed, with
use of the SPH method, spiral shocks on the assumption that the
specific heat ratio $ \gamma $ was 1.2.  Lanzafame \& Belvedere,
\cite{rf:25,rf:26} Boffin, Haraguchi \& Matsuda\cite{rf:27} 
also observed spiral
shocks in their recent SPH calculations.  This disagreement is
considered to be due to the difference in resolution, more concretely,
in the number of particles.  That is, use of a sufficient number of
particles, which enhances the resolution, leads to the observation of
spiral shocks.

With the finite difference/volume method, having a high accuracy
compared to the SPH method, it is expected that a detailed structure
of an accretion disk be readily observed.  Sawada \& Matsuda\cite{rf:28}
performed simulation by means of the three-dimensional finite volume
method for the first time.  Although the result shows the presence of
spiral shocks, no definite statement was made with respect to this
phenomenon, since the calculation time adopted by them was as short as
half an orbital period.

In recent years, our group conducted a series of the three-dimensional
finite volume calculations and confirmed the presence of spiral shocks
in all cases.\cite{rf:29,rf:30,rf:31} \ These calculations, however,
restricted the calculation region to the vicinity of the compact star
and did not include the companion.  The gas was assumed to flow into
through a rectangular hole located at the Lagrangian L1 point. Since
most of the calculation region was filled with a low-pressure gas, the
gas flowing into the region from the hole underwent a rapid expansion,
thereby forming what is known as under-expanded jet.  The flow (L1
stream) thus formed appeared to interact with the accretion disk in a
complex manner.  The hole, being artificial, made the L1 stream an
under-expanded jet having a square cross section, which was unnatural.
It has therefore been desired that simulation be made with the
calculation region extending up to the companion, whereby the true
shape of the L1 stream could be observed.

Bisikalo et al.\cite{rf:32}\tocite{rf:36} also made very similar works
to the present work, and the relation between our results and theirs
will be discussed in the last section.

\subsection{Difference between 2D and 3D Results}

Although radiative cooling has an important effect on the behavior of
an accretion disk, it is however very difficult to incorporate the
effect accurately in three-dimensional calculations.  To avoid this
difficulty, we use as the equation of state of the gas the one for an
ideal gas and, as the specific heat ratio of the gas $\gamma$, a value
smaller than 5/3, thereby simulating the effect of radiative cooling
to some extent.  

\begin{figure}
\epsfxsize=12cm
\centerline{\epsfbox{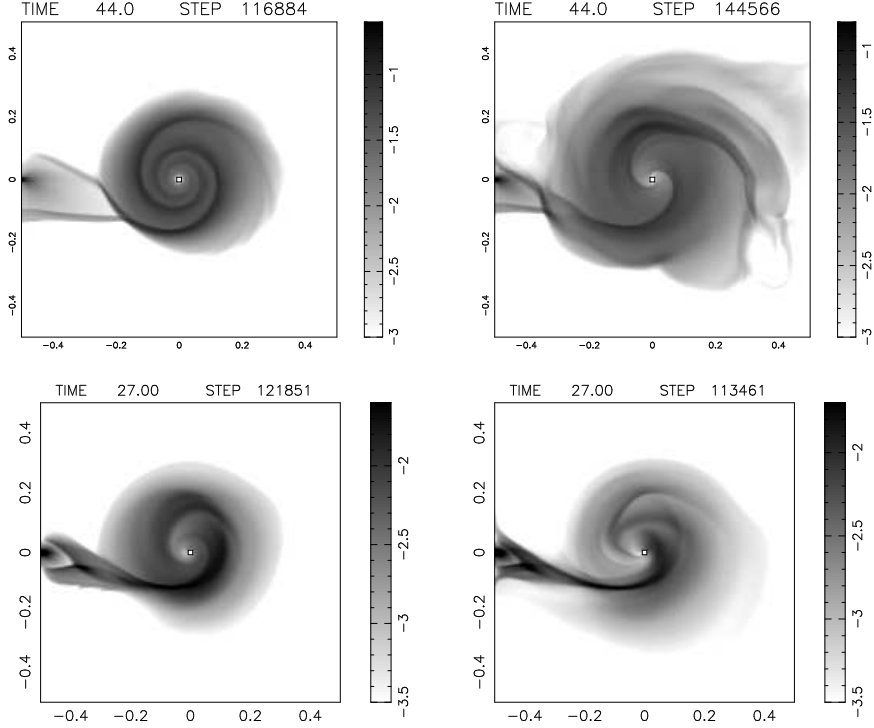}}
\caption{Difference between the 2D and 3D results: The density 
distributions obtained by 2D calculation (top) and those by 3D 
calculation on the rotational plane (bottom), with the specific heat 
ratio $\gamma$ being 1.01 (left) and 1.2 (right).  The calculation 
region is restricted to the vicinity of the compact star and exclude 
the companion. With 2D calculations smaller $\gamma$ leads to
tighter winding. On the other hand, 3D calculations do not show such
an effect clearly. The strange shape of the L1 stream is due to the
fact that the L1 stream comes from a square hole placed at the L1
point and, as a result, forms an under-expanded jet. (After Makita et
al.\cite{rf:31})}
\label{fig:1}
\end{figure}

What has puzzled us most in interpreting the results obtained by our
three-dimensional calculations is that they greatly differ from those
obtained by two-dimensional calculations (see figure \ref{fig:1}). 
With the two-dimensional calculations, we can obtain a clear correlation
between $ \gamma $ and the degree of winding-in of spiral shocks:
smaller pitch angle and tighter winding angle with smaller $ \gamma
$ (Makita et al.\cite{rf:31}, Matsuda et al.\cite{rf:37}).  
A smaller $\gamma $
corresponds, with the gas undergoing an adiabatic change, a smaller
temperature elevation.  The limit of $\gamma = 1 $ corresponds to an
isothermal change.  In the presence of shocks this is not necessarily
so, but with two-dimensional calculations, a small $ \gamma $
generally means a low temperature of the accretion disk, which
corresponds to a low sound speed and thus to a large Mach number.  It
is natural that in this case the spiral shocks wind in tightly.

On the other hand, as is clear from figure 1, with three-dimensional
calculations, the obtained spiral shocks seem to wind in very mildly,
in particular when $\gamma = 1.01 $.

This finding has long been a mystery to us.  Two possible
explanations may be given for this difference: numerical and physical
ones. Numerical explanations are:
\begin{enumerate}
\item The calculation region does not contain the companion, so that
the L1 stream does not have a natural shape. This fact may affect the
structure of the accretion disk.

\item In early stages of the calculation, a high-temperature gas is
assumed to be present over the entire calculation region.  This
assumption is necessary for the calculation to proceed stably.  In
two-dimensional calculations, an accretion disk forms with elapse of
time, while the gas initially placed is removed from the region of the
accretion disk and has no influence on the result.

\item With three-dimensional calculations, however, the
high-temperature gas remains above the accretion disk and may mix with
the disk gas, thereby increasing the disk temperature.  This effect
may be significant at the central part of the disk where it is thin.
\end{enumerate}

In order to solve these problems, the present study enlarges the
calculation region to an extent covering the companion.  Besides, in
order to check the influence of the high-temperature gas initially
placed, the temperature and density of the gas are intentionally
decreased.  As a result it has been found that the above numerical
factors are not crucial in causing the difference between the
calculation results.  It is then concluded that the difference in the
results between two-dimensional and three-dimensional calculations
should be explained on physical basis rather than numerical basis.

The results of the present calculations taking the companion into
account, to be shown later, indicate that the basic features of the
accretion disk does not differ very much from that obtained by
calculations without the companion.

\subsection{Penetration of the L1 Stream into the Disk and the Loss of
the Angular Momentum of the Disk Due to the Penetration}

Our calculations covering a region including the companion has
revealed that the L1 stream, on collision with the disk, penetrates
into the disk like a spear and does not decrease its velocity by
forming a hot spot.  We refer to this phenomenon as ``penetration''
hereinafter.  The presence of the penetration was not convincingly
confirmed in the previous three-dimensional calculations due to the L1
stream having an unnatural shape.  However, here, thanks to a
visualization technique adopted by us, the L1 stream reveals its
detailed structure. Note that {\it with two-dimensional calculations
the penetration can never occur}, which is understandable from
geometrical consideration.  This is an essential difference between
three-dimensional and two-dimensional calculations.

\begin{figure}
\epsfxsize=12cm
\centerline{\epsfbox{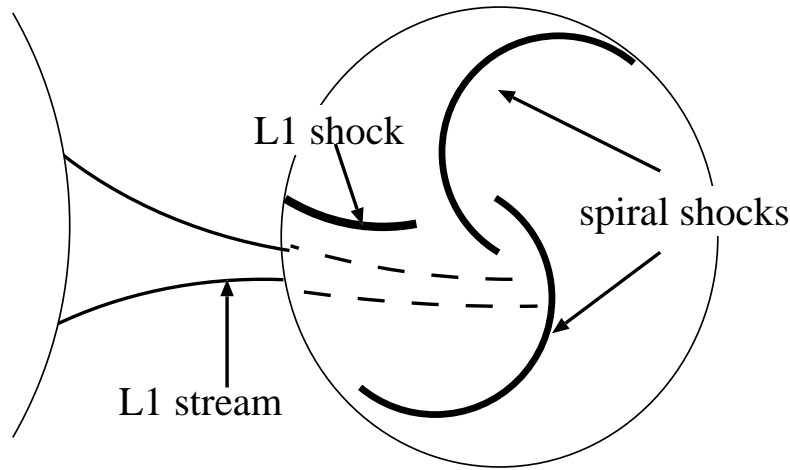}}
\caption{Schematic presentation of the flows in the accretion disk,
showing: the companion star (left end), the accretion disk (a 
large circle), the spiral shocks, the L1 stream flowing from the 
companion to the accretion disk through L1 point and the L1 shock, on 
which rotating disk gas collides with the L1 stream and deflects its 
direction.}
\label{fig:2}
\end{figure}

The disk gas supersonically rotating around the compact star must
form, on collision with
the high-density, low-temperature L1 stream resembling a spear, a bow
shock, which we call L1 shock. The bow shock thus formed
may be viewed as a hot line or a heated wall rather than a hot spot.
At the L1 shock the disk gas gives the penetrating L1 stream 
its angular momentum. The gas is at last released into the inner parts 
of the disk and mixes with the disk gas. The angular momentum, which is
transferred from the disk gas to the penetrating L1 stream at the L1
shock, is therefore finally redistributed into the disk, if tidal
torque from the secondary star dose not work.
But, the L1 shock and the penetrating L1 stream enhance the
non-axisymmetry in density distribution produced by spiral shocks in
the disk and, as a result, tidal torque from the secondary star transfers
more effectively the angular momentum of the disk to the angular
momentum of the orbital rotation of the binary system. The disk gas
therefore loses angular momentum and accretes onto the primary star.
In figure \ref{fig:2} schematic presentation of the spiral shocks, the 
L1 stream and the L1 shock are shown.

Here we present an analogy between a gas centrifuge to enrich uranium
and the accretion disk with the L1 stream. In order to enrich uranium,
there have been available ultra gas centrifuges, which comprise a
cylinder rotating at a high speed and containing uranium
hexa-fluoride.  The cylinder is equipped in the inside thereof a
stationary tube named ``scoop'' for recovering enriched or depleted
products of the uranium compound.  The rotating gas forms a bow shock
on collision with the scoop, and loses its angular momentum, whereby a
meridional flow called counter current is formed (Matsuda, Tamura \&
Sawada\cite{rf:38}). The scoop works as a sink of 
angular momentum, because the scoop and gas centrifuge are fixed to ground.
The L1 stream in the present case however dose not work as a sink of
angular momentum itself. The role of the L1 stream is only to get
angular momentum from the disk gas, and the secondary star actually
plays the role of a sink of angular momentum.

When calculation is made by means of a three-dimensional simulation,
the angular momentum of the disk gas is lost on collision with the L1
stream and, at the same time, the entropy and temperature of the disk
gas increase.  As a result, the sound speed increases and the Mach
number in the disk is reduced compared to that with two-dimensional
calculations, so that the spiral shocks wind in mildly.

Section 2 briefly describes the method of calculation.  Section 3
discusses the effect of the numerical assumptions on the results.
Section 4 discusses the detailed structure of the accretion disk and
the L1 stream. We compare the three-dimensional calculations and
two-dimensional ones in detail and shows that the difference can be
explained on a physical basis rather than on a numerical one.  Section 5
gives a summary and discussion.

\section{Numerical Models}

\subsection{Method of Numerical Simulation and Basic Assumptions}
We use, as before, the Simplified Flux Vector Splitting (SFS) finite
volume method proposed by Jyounouchi et al.,\cite{rf:39} Shima \&
Jyounouchi.\cite{rf:40} \ This method, being a variation of AUSM type
methods, has the feature of being simple and stable. As to the detail
of the numerical scheme and the SFS scheme see 
Makita et al.\cite{rf:31} \ 
We use a MUSCL-type technique, thereby keeping the spatial and
temporal accuracy at second order levels.  We consider the gas as a
perfect gas and do not take into account such complex effects as
viscosity, radiation and magnetic field.  The equation of state is as
follows.
\begin{equation}
\label{eq:1}
p=(\gamma-1)\left[e-\frac{1}{2}\rho(u^2+v^2+w^2)\right],
\end{equation}
where $p,\rho, e$ are the pressure, density and total energy per unit
volume of the gas, respectively, and $\gamma$ is the specific heat
ratio; $u, v,w$ are the velocity component in the directions of
$x,y,z$, respectively.  We use as the specific heat ratio, not the
value of 5/3, which is that of an adiabatic gas, but smaller values
and mainly $\gamma = 1.01, 1.2$.  In the absence of shocks, the
entropy would be kept constant and the temperature would change as $T
\propto \rho^{\gamma-1}$.  The temperature change thus becomes nearly
isothermal at a limit of $\gamma$ approaching 1.  Our calculations,
however, produce shocks, where the entropy of the gas increases.
Accordingly, $ \gamma \sim 1$ does not necessarily mean that the
accretion disk is nearly isothermal.

Indeed, with use of the same value for $\gamma$, the temperature
and the entropy of an accretion disk obtained by means of
three-dimensional calculations differ from those by two-dimensional
ones significantly. This is attributable to whether or not shocks are
present and to the difference in the strengths of the shocks.  By
selection of $\gamma = 1.01$, we intend to increase the cooling
effect as much as possible while keeping numerical accuracy.  The
larger $\gamma = 1.2$ is selected because a $\gamma$ larger than
this would not produce an accretion disk itself.  We thus study the
two extreme cases.

We consider, as the model of a binary, a system consisting of a
compact star (primary star) with mass $M_1$ and a mass-losing
companion with mass $M_2$.  The mass ratio $q=M_2/M_1$ is fixed at 1
in the present study, while with IP Peg $q \sim 0.5$.  As to other
mass ratios see the paper by Matsuda et al.,\cite{rf:30} although they
restricted their computational region to the vicinity of the primary
star.

All physical quantities are normalized.  The length is scaled by the
separation $A$.  With the rotational frequency of the binary being
$\Omega$, the time is scaled as $1/\Omega$, so that the orbital period
becomes $2 \pi$.  The density at the surface of the companion is taken
to be 1.  The gravity constant $G$ is eliminated through the following
equation.

\begin{equation}
\label{eq:2}
A^3 \Omega^2=G(M_1+M_2).
\end{equation}

A Cartesian coordinate system with the origin located at the center of
the primary star is used.  The x-axis is a line combining the centers
of the primary and the companion.  The center of the companion is at
$(-1,0,0)$. The z-axis is perpendicular to the orbital plane and is
oriented in the same direction as the angular momentum vector of
orbital rotation.  The $x-y$ plane therefore constitutes the orbital
plane.

The calculation region is a rectangular parallelepiped covering from
$(-1.5,-0.5,0)$ to $(0.5,0.5,0.5)$.  With the assumption of symmetry
of physical quantities in relation to the orbital plane, the
calculations are made only in the region of upper half.  This region
is divided with grid points of $201 \times 101 \times 51$.

\subsection{Boundary Conditions and Initial Conditions}
The primary star is assumed to be a hole occupying a region covering
$3 \times 3 \times 2$ and centered at the origin.  The inside of the
primary star is always filled with a gas having a density, sound speed
and velocity of $\rho_1$, $c_1$ and 0, respectively. The pressure in
the inside of the hole is kept sufficiently low in order to always
absorb the surrounding gas.  Here $\rho_1$ and $c_1$ are parameters.
The mass accretion rate onto the main star is obtained by solving
Riemann problem at each time step.

The companion is assumed to fill the critical Roche lobe.  The inside
of the companion is filled with a gas with density $\rho_0=1$, sound
speed $c_0$ and velocity 0, while the gravity is neglected in the
companion. Here $c_0$ is taken as a parameter and its effect is
studied.  The assumption that the gas inside the companion has 0
velocity does not mean that there is no gas outflow from the
companion. The gas does flow out when the pressure outside the
companion is lower than that inside the companion.  The outflow flux
is calculated by solving Riemann problem at each time step.

The outside of the outer boundary is assumed to be always filled with
a gas having a density $\rho_1$, sound speed $c_1$ and velocity 0.
Use of this boundary condition insures stable calculation.  Here also
inflow or outflow of the gas through the outer boundary is possible
and the flux is calculated by solving Riemann problem between the
inside and outside of the calculation region at each time step.

As the initial conditions, the whole region except the inside of the
companion is occupied at time $t=0$ by the gas with density $\rho_1$,
sound speed $c_1$ and velocity 0.  The initial density $\rho_1$ and
initial sound speed $c_1$ are parameters, the effects of which are
studied later. The model parameters used in the present work are
summarized in the table 1.

\begin{table}[t]
\caption{Parameters.}
\label{table:1}
\begin{center}
\begin{tabular}{cccc} \hline \hline
Model & $c_0$ & $\rho_1, c_1$ & $\gamma$ \\ 
\hline \\
A & $0.02$ & $10^{-5}, \sqrt{10}$ & $1.01$  \\
B & $0.02$ & $10^{-5}, \sqrt{10}$ & $1.2 $  \\

C & $0.02$ & $10^{-7},      0.02$ & $1.01$  \\
D & $0.02$ & $10^{-7},      0.02$ & $1.2 $  \\

E & $0.1 $ & $10^{-5}, \sqrt{10}$ & $1.01$  \\
F & $0.1 $ & $10^{-5}, \sqrt{10}$ & $1.2 $  \\

\hline
\end{tabular}
\end{center}
\end{table}

Typically, calculations are continued up to an orbital period of more
than 10.  We confirm that the disk mass does not change appreciably
after a few rotation period, so we may say that our computational
results are nearly in steady state except insignificant oscillations.

\section{Effect of Various Parameters on the Results of Calculations}
\subsection{Effect of the Conditions of Gas Placed Initially in the
Calculation Region}
In our previous calculations,\cite{rf:29,rf:30,rf:31}
a gas with a density $\rho_1=10^{-5}$
and a sound speed $c_1=\sqrt{10}$ was placed at time $t=0$.  This high
sound speed means that the gas has a high temperature.  This selection
was made, since it was considered that such a high-temperature,
low-density gas hardly falls onto the accretion disk or, if falls at
all, affects the accretion disk only to a small extent.
  
The results obtained by the three-dimensional calculations, however,
differed greatly, as described in Introduction, from those by the
two-dimensional calculations.  That is, the spiral arms with the
three-dimensional calculations wind in more mildly than those with the
two-dimensional calculations, even with the specific heat ratio
$\gamma=1.01$.  This is attributable to the high temperature of the
accretion disk obtained in the three-dimensional calculations, which
in turn may be due to the high-temperature gas initially placed.

Unlike with two-dimensional calculations, the gas initially placed
will, during progress of calculation and after an accretion disk has
been formed, remain above the accretion disk. This initial gas may
mix with the accretion disk, thereby increasing the temperature. We
selected the density of the initial gas to be $10^{-5}$, which we
considered to be sufficiently small. However, the central part of the
accretion disk, having a small thickness, may have been affected.

\begin{figure}[htb]
\hspace{1cm}
   \parbox{6cm}{
      \epsfxsize=5.5cm
      \epsfbox{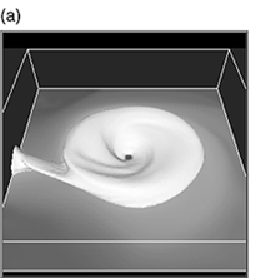}
   }
   \hspace{8mm}
   \parbox{6cm}{
      \epsfxsize=5.5cm
      \epsfbox{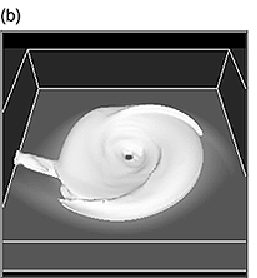}
   }
\caption{Effect of the initially placed gas: Iso-density surfaces at 
$\log \rho=-3.3$ of the disk for Model A with high temperature initial 
gas (a) and that at $\log\rho=-4.0$ for Model C with low
temperature and low density initial gas (b). The Model C shows
clearer spiral shocks. In both cases, spiral shocks wind
more loosely than that in two-dimensional calculations with
$\gamma=1.01$.}
\label{fig:3}
\end{figure}

In the present study, we compare two cases with the initial gas having
different densities and temperatures, in order to check their
influence. Figure \ref{fig:3} shows the results. We may observe that 
Model C with low temperature and low density initial gas 
shows clearer and more openly wind spiral shocks than Model
A with high temperature and high density initial gas. 
However, the difference may not be very significant,
and in both cases, spiral shocks wind loosely even with $\gamma=1.01$.

The key role is played, not by the temperature itself, but by the heat
capacity, of the initial gas.  A sufficiently low density of the
initial gas therefore causes no problem with respect to heat capacity
even when it has high temperature.  Therefore, we may conclude that
our previous choice of the initial condition is good enough, as far as
the cases of $\gamma= 1.01$ are concerned. The cases of $\gamma=1.2$
are slightly affected by the initial condition as will be discussed in
3.4.

\subsection{Effect of the Surface Temperature of Companion}
In our previous calculations, the gas placed at L1 point was assumed
to have a sound speed of $c_0=0.1$.  The present study includes the
case of the gas present on the companion having a sound speed of
$c_0=0.1$, which is, however, very large in actuality. If the binary
has a typical speed of 500 km s$^{-1}$, the sound speed becomes 
50 km s$^{-1}$, corresponding to a temperature of as high as 
$2.5 \times 10^{5}$ K.

This high sound speed has been selected, because a high temperature on
the surface of the companion increases the rate of the gas flowing out
through the L1 point, so that calculation can proceed fast. Since,
however, the above high temperature is unrealistic, the present study
also includes the case of $c_0=0.02$ corresponding to $10^4$ K.

\begin{figure}[htb]
\hspace{1cm}
   \parbox{6cm}{
      \epsfxsize=5.5cm
      \epsfbox{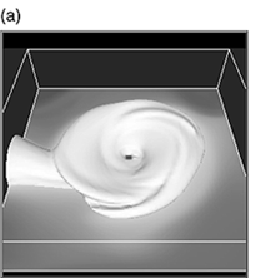}
   }
   \hspace{8mm}
   \parbox{6cm}{
      \epsfxsize=5.5cm
      \epsfbox{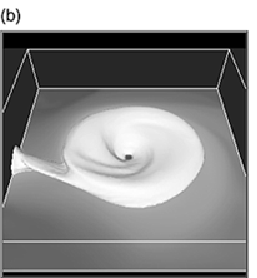}
   }
\caption{Effect of the surface temperature of companion: Iso-density 
surfaces at $\log\rho=-2.5$ for Model A with high surface temperature 
of companion (a), and that at $\log\rho=-3.3$ for Model E with low surface
temperature (b). With the high surface 
temperature of companion, the L1 stream becomes thick and the 
inflow rate high and the disk becomes large. In Model E, 
spiral shocks become clearer and wind more loosely than 
that in Model A.}
\label{fig:4}
\end{figure}

Figure \ref{fig:4} shows, with $\gamma=1.01$, the influence of the 
surface temperature of the companion.  With the high temperature, the 
L1 stream becomes thick and its inflow rate increases.  
On the other hand, with the low temperature, which may be more 
realistic, the L1 stream becomes thin. Model E shows clearer 
and more widely open spiral shocks than Model A, because of 
larger disk size.

\subsection{Effect of Specific Heat Ratio}
As commented in Introduction, the influence of radiative cooling is
very important in accretion disks.  Without this cooling, the
temperature of an accretion disk would become the virial temperature,
which is very high.  It is however not practical to perform any
three-dimensional simulation of radiative fluid dynamics accurately
with computing capability currently available. 

Therefore, workers in this field have been used either an eqation
of state with low specific heat ratio like us, that of iso-thermal 
gas or that of iso-entropy gas.
A low specific heat ratio is selected in the present work in order to 
emulate the influence of radiative cooling. Although the ionized plasma 
gas present in actual accretion disks has a specific heat ratio of 5/3, 
the present calculations use $\gamma=1.01, 1.2$.  Our calculations thus, 
in a strict sense, simulate a poly-atomic molecular gas. In this case, 
when the gas is compressed, its energy, being distributed to the degree of
internal freedom, will not contribute to translational motion to a
large extent, so that the temperature elevation is suppressed.
This internal energy is comvected with gas and eventually accreted by
the central object, and is not radiated away to the outer space.
Nevertheless the low ratio of specific heats emurates radiative cooling
somehow. The caluculation including the radiative transfer is now 
work in progress (Nagae et al. 2001).

\begin{figure}[htb]
   \parbox{4.5cm}{
      \epsfxsize=4.5cm
      \epsfbox{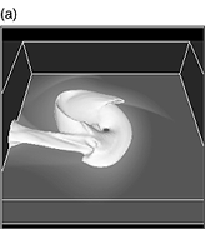}
   }
   \hspace{2mm}
   \parbox{4.5cm}{
      \epsfxsize=4.5cm
      \epsfbox{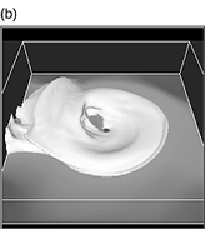}
   }
   \hspace{2mm}
   \parbox{4.5cm}{
      \epsfxsize=4.5cm
      \epsfbox{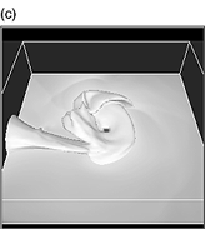}
   }
\caption{Iso-density surfaces for a specific heat ratio of 
$\gamma =1.2$, under different boundary conditions and initial 
conditions: (a) Model B at $\log\rho =-3.7$. (b) Model D at 
$\log\rho =-4.0$, with low density and low temperature of initial gas. 
(c) Model F at $\log\rho =-2.9$, with high temperature at the surface
of the companion. With $\gamma=1.2$, disk is affected significantly by 
both initial and boundary conditions.}
\label{fig:5}
\end{figure}

In the limit of $\gamma=1$, the gas becomes isothermal.  The equation
of state adopted by us, however, does not allow $\gamma=1$.  We
therefore select $\gamma=1.01$, which is the minimum value in view of
calculation accuracy. As a counterpart, we select $\gamma=1.2$, above
which experience tells us no accretion disk can ever form.  Figure
\ref{fig:5} shows the results.

With $\gamma=1.01$, a clear accretion disk and shock waves appear
under all the initial conditions or boundary conditions assumed.  On
the other hand, with $\gamma=1.2$, the shape of the accretion disk
finally obtained varies depending on the initial conditions and
boundary conditions.  In Model B and F the accretion disks are highly
distorted, while in Model D we may observe a normal accretion disk.
In all cases, the temperature of the disk is more higher than that
of the disk in its outburst.

Bisikalo et al.\cite{rf:32}\tocite{rf:36} argue that no accretion disk 
forms with $\gamma=1.2$, which is not completely inconsistent with 
our results.  At any rate, it is expected that a specific heat ratio
larger than 1.2 may produce no accretion disk under any initial or
boundary conditions.

\section{Detailed Structure of Flow}
\subsection{Penetration of the L1 Stream into the Accretion Disk and
Formation of an L1 Shock}
In order to investigate the detailed structure of the accretion disk
and the L1 stream, let us investigate the details of some specific
models in this section. Since the case of specific heat ratio
$\gamma=1.2$ corresponds to too high a temperature of the disk, only
the case $\gamma=1.01$ is studied here.  In the present case,
$c_0=0.02$ is taken as the sound speed on the surface of the
companion.

\begin{figure}[htb]
\hspace{1cm}
   \parbox{7cm}{
      \epsfxsize=5.5cm
      \epsfbox{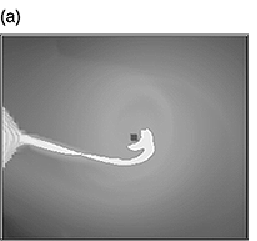}
      \epsfxsize=5.5cm
      \epsfbox{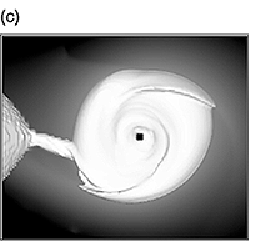}
   }
   \hspace{5mm}
   \parbox{7cm}{
      \epsfxsize=5.5cm
      \epsfbox{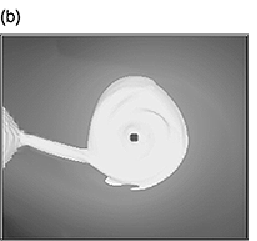}
      \epsfxsize=5.5cm
      \epsfbox{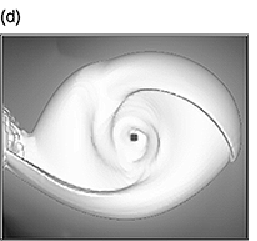}
   }
\caption{Iso-density surfaces for Model C at various density levels: (a)
$\log\rho=-2.0$. (b) $\log\rho=-3.0$. (c) $\log\rho=-4.0$.
(d) $\log\rho=-5.0$.}
\label{fig:6}
\end{figure}

\begin{figure}
\epsfxsize=8cm
\centerline{\epsfbox{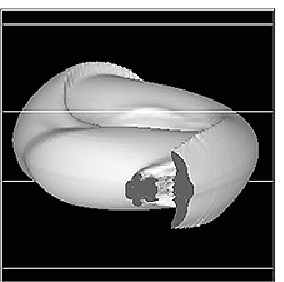}}
\caption{The same as figure \ref{fig:6} at $\log\rho = -4.0$ 
but seen from a different angle. One may observe cross sections of a 
spiral shock and the L1 shock. The spiral shock is concave to the
upstream, while the L1 shock is convex to the upstream. This means
the L1 stream is a bow shock.}
\label{fig:7}
\end{figure}

Figure \ref{fig:6} shows the iso-density surfaces at various 
levels of Model C. And figure \ref{fig:7} shows the iso-density 
surface of the same model from a different view angle. 
One may observe cross-sectional structures of a
spiral shock and the L1 shock. The spiral shock seems to be concave
to the upstream, while the L1 shock is convex. One may conclude that
the L1 shock is a bow shock.

The figures so far presented show only iso-density surfaces, which
cannot reveal any detailed structure, in particular the velocity
distribution, of the flow.  Hereafter let us investigate the
interaction between the L1 stream and the accretion disk, with use of
the velocity distribution.

\begin{figure}
\epsfxsize=5cm
\centerline{\epsfbox{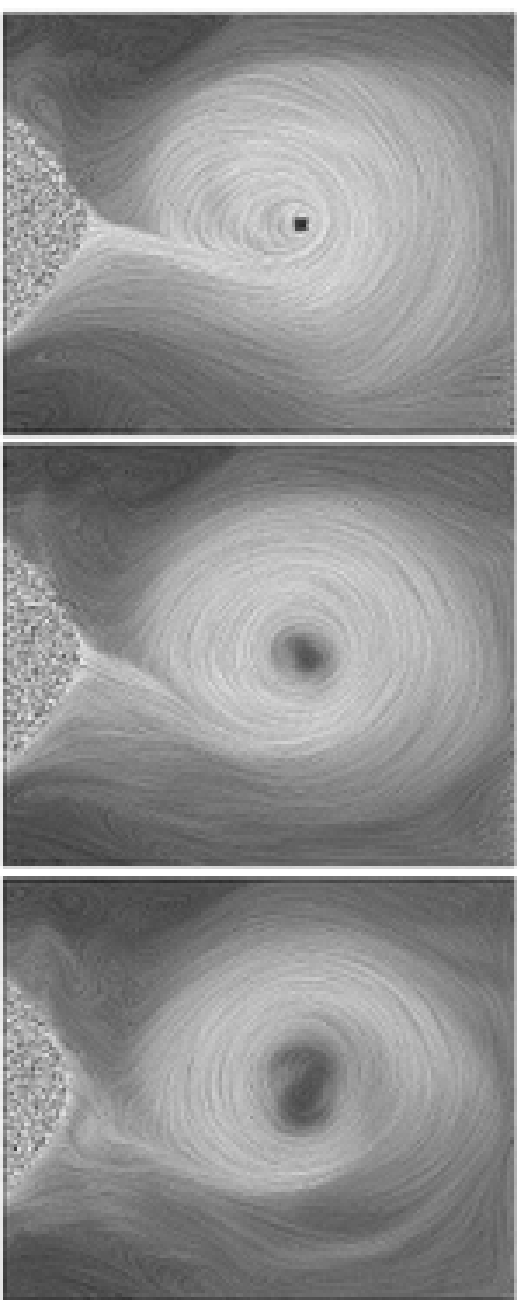}}
\caption{Flow patterns on the orbital plane and on x-y planes at various
heights, for Model A, are shown by means of the LIC method. The 
degree of whiteness represents the magnitude of density.
From top to bottom $z=0$, $z=0.04$; and $z=0.08$.  
On the orbital plane with $z=0$, the gas in the accretion disk 
collides with the L1 stream and, as a result, loses its angular 
momentum and falls towards the central star.  On $z=0.08$, the gas 
of the accretion disk rotates about the central star.  The 
influence of the L1 stream is thus limited to a region near the 
orbital plane.}
\label{fig:8}
\end{figure}

Figure \ref{fig:8} shows velocity distributions on x-y planes at 
various heights by means of the Line Integration Convolution (LIC) method.
The figure shows that on the
orbital plane of $z=0$, gas in the accretion disk collides with the L1
stream and, as a result, loses its angular momentum and falls towards
the primary star.  A bow shock is then formed close to and parallel
with the high-density, spear-like L1 stream (above the L1 stream in
the figure). On $z=0.08$, the gas of the accretion disk rotates about
the central star.  The influence of the L1 stream is thus limited to a
region near the orbital plane.

\begin{figure}
\epsfxsize=5cm
\centerline{\epsfbox{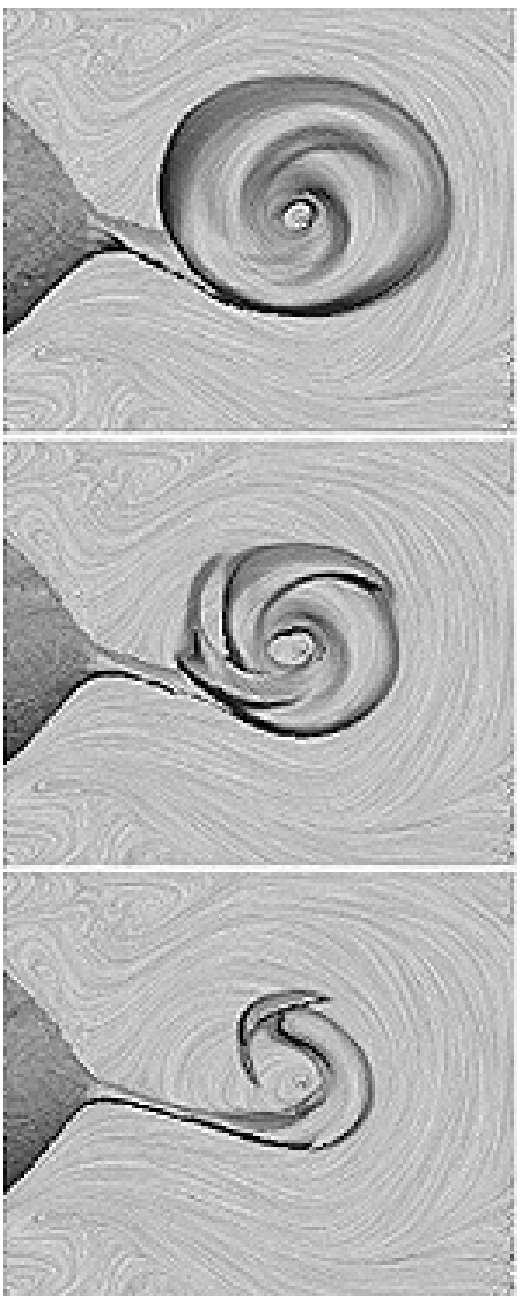}}
\caption{Iso-density surfaces and flow patterns thereon, together 
with a flow pattern on the orbital plane, of Model A drawn by 
LIC method. From top
to bottom at $\log \rho=-3.3, -3.0, -2.7$ all at time $t=75.36$.}
\label{fig:9}
\end{figure}

\begin{figure}
\epsfxsize=9cm
\centerline{\epsfbox{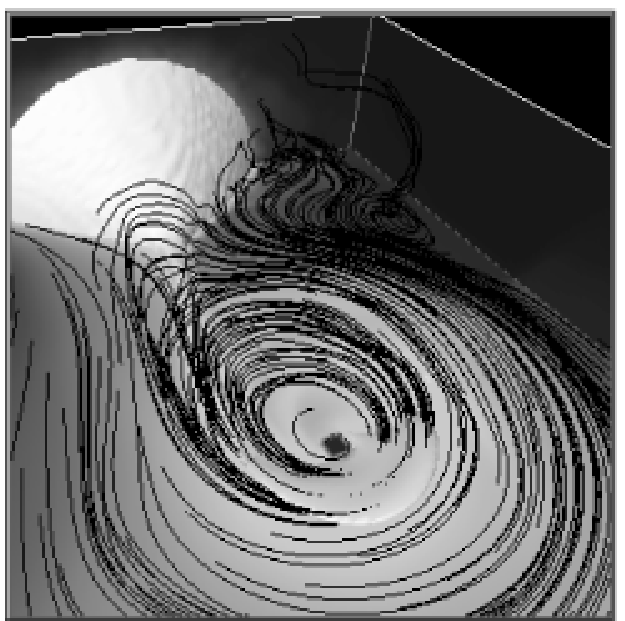}}
\caption{Three-dimensional view of streamlines of model A.}
\label{fig:10}
\end{figure}

Figure \ref{fig:9} presents, in order to observe the above phenomena 
more closely, flow patterns, drawn by means of the LIC method, on the
iso-density surface.  The figure shows the internal structure more
clearly with increasing density of the iso-density surface. Figure
\ref{fig:10} shows a three-dimensional view of the streamlines.

These pictures are particularly interesting, because they show two
remarkable features. 
1) The L1 stream penetrates into the accretion disk.  
2) The disk gas having collided with the L1 stream abruptly
changes, due to the L1 bow shock formed, its direction towards the
primary star.

\subsection{Influence of the L1 Stream on the Disk Gas}
\begin{figure}[htb]
   \parbox{6cm}{
      \epsfbox{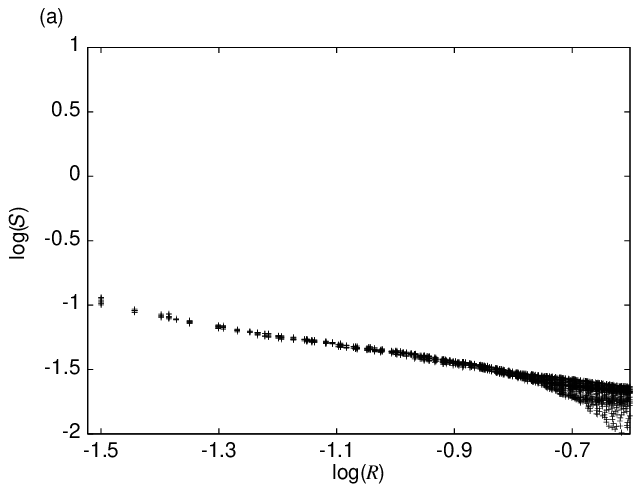}
      \epsfbox{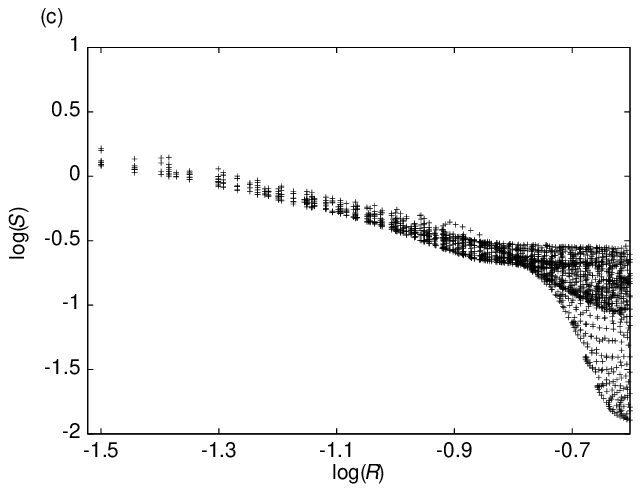}
    }
  \parbox{6cm}{
      \epsfbox{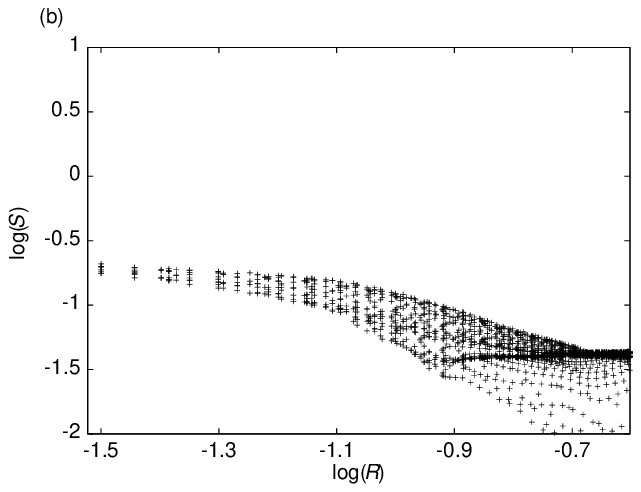}
      \epsfbox{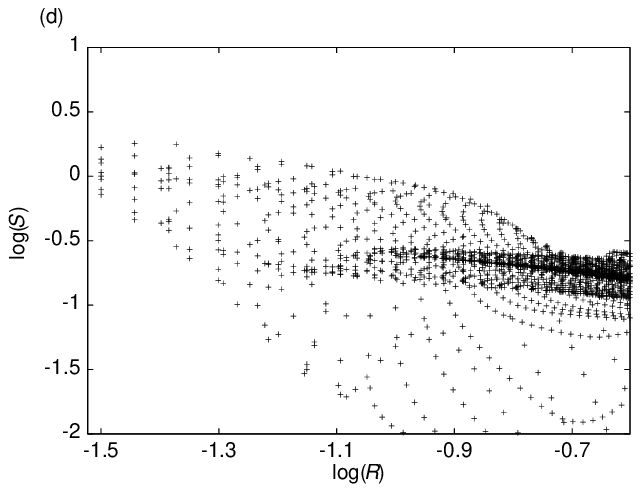}
   }
\caption{Radial distribution of entropy $\log(p/\rho^{\gamma})$ in the
accretion disk.  The top row corresponds to $\gamma=1.01$, and the
bottom row to $\gamma=1.2$.  The left-side column shows the case with
two-dimensional calculations and the right-side column that with
three-dimensional calculations.  Other parameters are $\rho_1=10^{-7},
c_1=0.02,$ and $c_0=0.02$. In three-dimensional calculations, entropy
increases rapidly at the peripheral zone due to the L1 shock. This
increase of entropy is significant with $\gamma=1.01$.}
\label{fig:11}
\end{figure}

Figure \ref{fig:11} shows, in order to study the influence of the 
L1 stream on
the disk gas, the distribution of the entropy related function of the
gas $S=p/\rho^{\gamma}$ obtained by two-dimensional and
three-dimensional calculations, with the abscissa representing the
distance from the central star and the ordinate the entropy:
$\log(S)$.  Dispersion in the plots in the ordinate direction
corresponds to that in the values at various angular positions.

We discuss, in particular, the case of $\gamma=1.01$, where, $S$
being almost the same as temperature, the figure can be considered to
be showing the temperature distribution.  The figure clearly shows a
smaller dispersion with two-dimensional calculations than with
three-dimensional calculations. This means that the entropy and
temperature distributions in two-dimensional cases are nearly
axisymmetric, which in turn means that the entropy increases due to
spiral shocks, only to a moderate extent.  The entropy values in the
central zone (the left end) and the peripheral zone (the right end)
are lower with two-dimensional calculations than with three-dimensional
ones.  A large dispersion in the values with three-dimensional
calculations is caused by the L1 shock generated on collision between
the L1 stream and the disk gas.  In particular, a horizontal linear
distribution of $S$, seen in a region corresponding to large radii, is
due to the L1 shock.  Accordingly, with three-dimensional
calculations, a high entropy or high temperature of the gas decreases
the Mach number of the gas, so that spiral shocks wind in mildly even
with $\gamma=1.01$.

Figure \ref{fig:11} also shows the entropy distribution for the case of
$\gamma=1.2$. In this case, although the distribution is wider in
three-dimensional calculations than two-dimensional ones, the absolute
values are almost the same in both calculations.  The angle of
winding-in of spiral shocks is therefore almost the same with both
two-dimensional and three-dimensional calculations.


\subsection{Effective $\alpha$}
The accretion disk has a large non-axisymmetry in its density 
distribution because of the loosely wound spiral shocks in our 
calculations. In the spiral shock model of angular momentum transfer, 
tidal torque from the secondary star transfers the angular momentum of 
gas in the accretion disk to the angular momentum of the orbital 
motion of the binary system. Angular momentum transfer becomes more 
effective as spiral shocks wind more loosely and, as a result, 
the non-axisymmetry of density distribution becomes large.

In our calculations, the disk gas loses its angular momentum 
not only through spiral shocks, but on interaction with the
penetrating L1 stream. 

The disk gas loses its angular momentum on collision with the 
penetrating L1 stream. In this process, angular momentum is 
transferred directly from the disk gas to the penetrating L1 
stream. This asngular momentum transfer is extremely 
effective on the equatorial plane. In figure 8, the streamlines of
the disk gas is bent on collision with the penetrating 
L1 stream. This mechanism is effective, but is closed
in the disk. The angular momentum transferred to the penetrating L1 stream
is finally released into the central part of the disk.
This mechanism is, therefore, angular momentum re-distribution rather
than angular momentum transfer.

The L1 shock produced on collision between
the disk gas and the penetrating L1 stream contributes to a 
non-axisymmetry in density distribution in the disk, and the disk 
gas loses its angular momentum through the L1 shock. 
These two mechanisms are effective in the peripheral part of the
disk into which the L1 stream is penetrating.

\begin{figure}
\epsfxsize=12cm
\centerline{\epsfbox{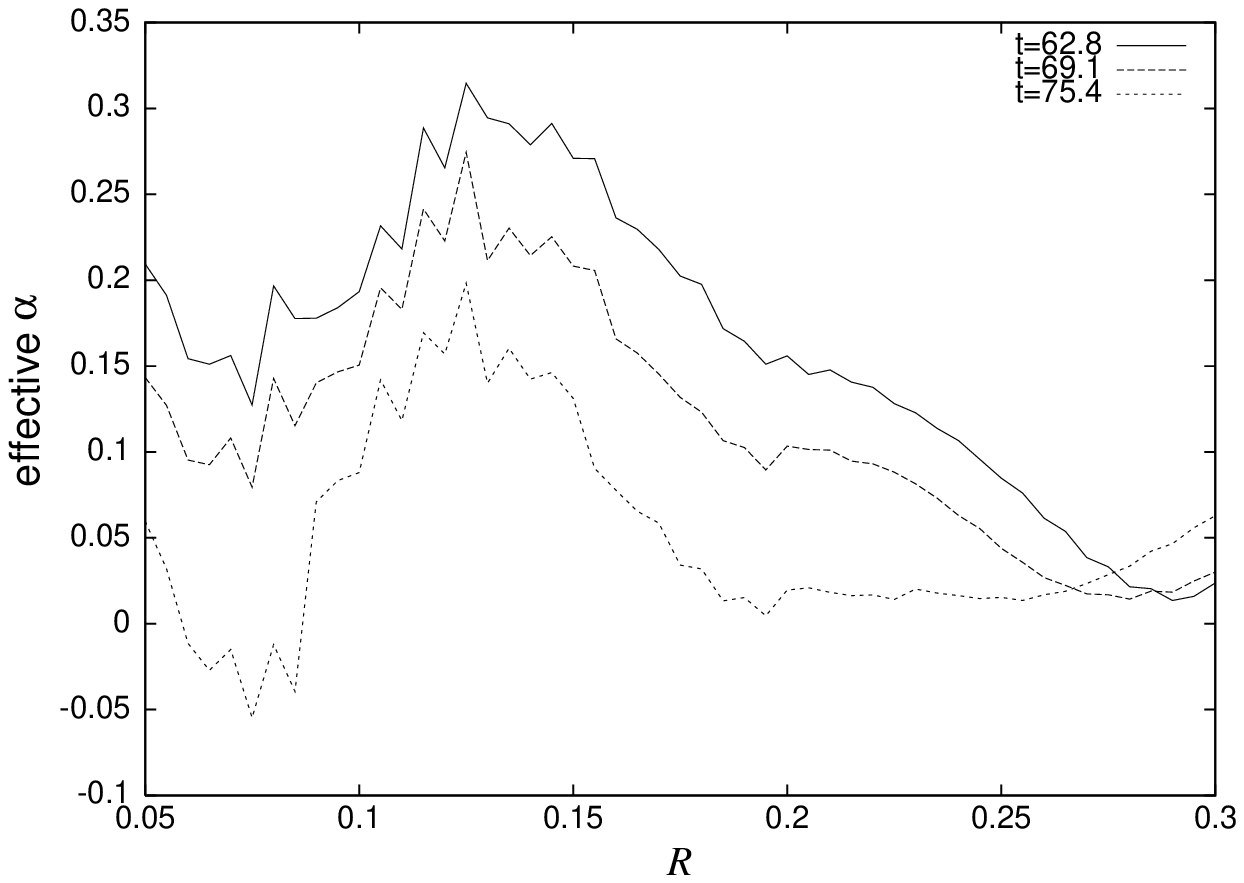}}
\caption{Radial distributions of effective $\alpha$ averaged in 
azimuthal direction at 10, 11 and 12 orbital periods for Model A. 
Effective $\alpha$ oscillates in time, but its peak is located at 
$R\sim1.2$ and the value decreases with radius. The maximum value of 
effective $\alpha$ is $\sim0.3$.}
\label{fig:12}
\end{figure}

We calculate the effective $\alpha$, according
to Frank, King \& Raine,\cite{rf:41}
\begin{equation}
\nu=\frac{\dot M}{3 \pi \Sigma}
\Bigr[1-\Bigr(\frac{R_{\ast}}{R}\Bigl)^{\frac{1}{2}}\Bigl],
\end{equation}
\begin{equation}
\alpha=\frac{\nu}{c_s H},
\end{equation}
where $\nu$ is kinematic viscosity, $\dot M$ is mass accretion rate,
$\Sigma$ is surface density, $R_{\ast}$ is radius of the primary
star, $R$ is distance from the primary star, $c_s$ is sound speed and
$H$ is scale height. We neglect
$\Bigr(R_{\ast}/R\Bigl)^{\frac{1}{2}}$ because
$R_{\ast}/{R}\ll 1$. Figure \ref{fig:12} shows the effective 
$\alpha$ at each orbital period for Model A. The effective 
$\alpha$ varies with both of radius and
time, its maximum being $\alpha \sim 0.3$. 

In Spruit\cite{rf:15}
and Livio \& Spruit,\cite{rf:42} the effective $\alpha$ depends 
on the mass ratio and the pitch angle of spiral shock. 
However, the analytical value of the effective $\alpha$ is of 
order of $\sim 10^{-2}$, while in two-dimensional calculations 
by Matsuda et al.\cite{rf:12}, the effective $\alpha$ is of 
the order $0.1$ with large $\gamma$. 

The effective $\alpha$ is large in the peripheral part. In this
region, the non-axisymmetric density distribution is enhanced by 
the L1 shock, and therefore, the disk gas loses
effectively its angula momentum due to tidal torque from the
secondary star. The disk gas also loses its angular momentum 
on collision with the penetrating L1 stream in the peripheral part.
On the other hand, in the central part, the effective $\alpha$ is
of the order $0.1$, and this value is similar to that in two
dimensional calculations with large $\gamma$ and loosely wound
spiral shocks.

\subsection{Formation of the 'hot spot'}
In our three-dimensional simulation, the effect of radiative cooling 
is neglected, and therefore the accretion disk becomes hot. This hot 
disk becomes thick and of low density, so that the dense L1 stream 
can penetrate into the disk and the 'hot line' is formed along the 
penetrating L1 stream.

On the other hand, in two-dimensional simulations, the disk cannot
expand along the $z$-axis, and therefore the density of the disk becomes
larger than that of the L1 stream. And the L1 stream collides 
with the disk edge and does not penetrate into the disk.\cite{rf:6}
\tocite{rf:14} \ Rozyczka \& Schwarzenberg-Czerny\cite{rf:43} and 
Rozyczka\cite{rf:44} performed two-dimensional simulations of the 
collision between the L1 stream and the disk edge. They showed that
the collision between the L1 stream and the disk edge forms two shock 
waves and concluded that these shock waves are considered as the 
'hot spot'.

The difference of the density between the disk and the L1 stream
is important.
Whether or not the L1 stream penetrates into the disk depends on the
magnitude of each of the densities. If the density
of the disk is larger than that of the L1 stream in the 
three-dimensional simulation, the L1 stream will be stopped on 
collision with the disk edge and the hot spot will be formed.

\begin{figure}
\epsfxsize=8cm
\centerline{\epsfbox{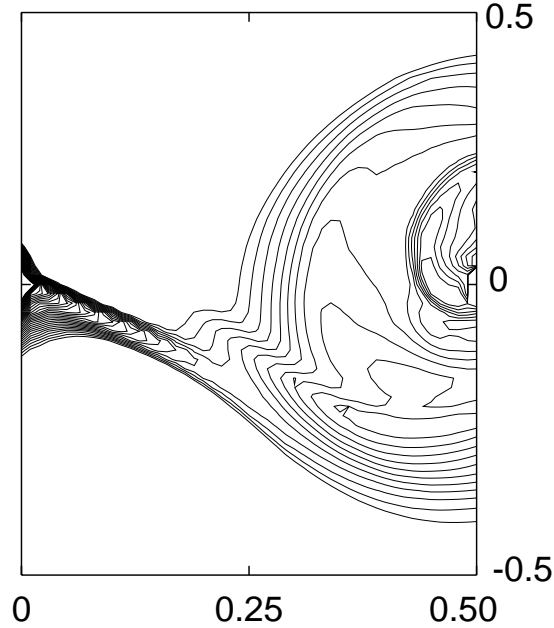}}
\caption{Density contours on the equatorial plane. The density of the
L1 stream at the collision point is of the same order of magnitude with the
density of the disk edge. The L1 stream is stopped on collision with the
disk edge and shock waves are formed. These features are similar 
to the results of two-dimensional simulations.}
\label{fig:13}
\end{figure}

In order to investigate collision between the L1 stream and the dense 
accretion disk, we suddenly changed the density of the surface of the 
secondary star from 1 to 0.5, after the accretion disk has reached 
steady state. In this case, the density of the L1 stream decreases 
and becomes of the same order of magnitude with the density at the disk edge.
Figure \ref{fig:13} shows the
density contours on the equatorial plane in the vicinity of the 
collision point. The L1 stream is stopped on collision with the disk 
edge, and shock waves are formed at the collision point. 
These features are similar to the results of two-dimensional simulations.

If the effect of radiative cooling is considered, the accretion disk
will becomes cool, thin and of high density. In such conditions, we consider
that the L1 stream cannot penetrate into the dense disk 
and that the 'hot spot' is formed on collision between the L1 stream 
and the disk. 

\subsection{Flows on the Companion}
We have so far paid attention to the accretion disk and the L1 stream. 
Since our calculation region includes the companion, it is also of
interest how the flows on the surface of the companion behave,
although we may not recognize this with observational means currently
available.

\begin{figure}
\epsfxsize=9cm
\centerline{\epsfbox{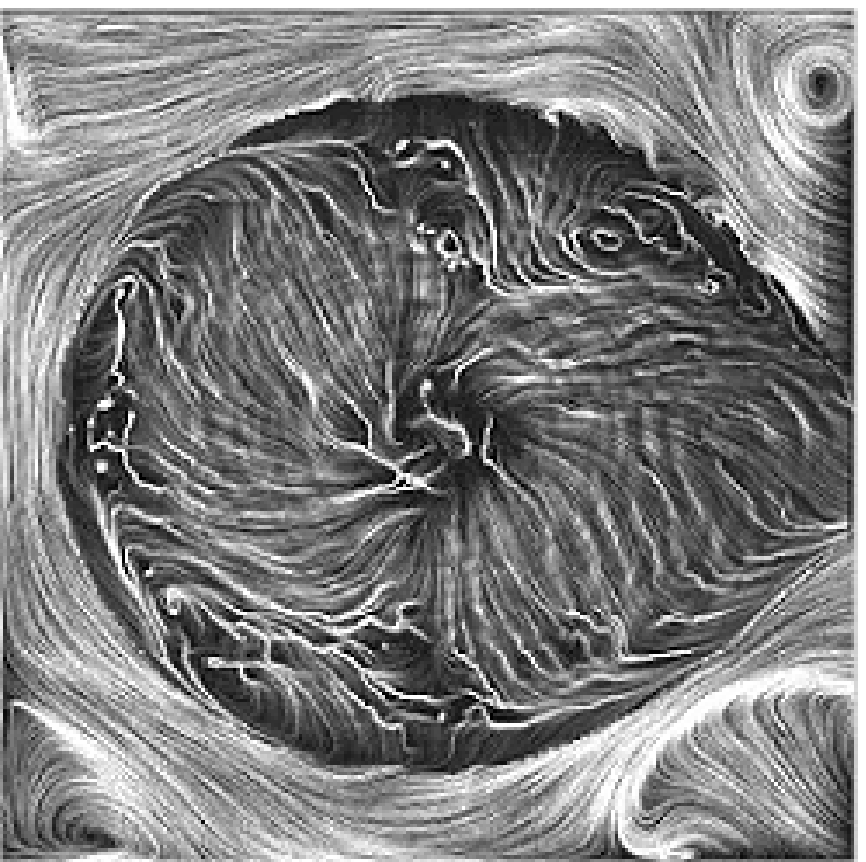}}
\caption{Flows on the surface of companion produced by Pixcel Exposure
Method for case A at $t=56.52$.
The value of density on which the flows are plotted is $\log 
\rho=-3.5$. Gas which outflows from high-latitude regions flows 
toward the L1 point. This main flow pattern is almost steady, 
although small flow patterns change.
Much attention should not be paid on vortices at the corner of the 
rotational plane, because they may be affected by the outer 
boundary condition.}
\label{fig:14}
\end{figure}

The coordinate system we employed is Cartesian, which can hardly
represent precisely the companion having a nearly spherical surface.
Outer boundaries may also affect the surface flow of the companion.
We have to examine the results with these restrictions in mind.
Figure \ref{fig:14} shows flows on the surface of the companion for 
case A at 9 orbital periods.
Figures are produced using Pixcel Exposure Method (PEM). 
Worth noticing is that flows towards the L1 point start even at 
high-latitude regions. This main flow pattern is almost steady,
although small flow patterns change.

It is also interesting to note that eddy patterns have been formed on
the surface of the companion.  This is considered to be similar to
vortices of low pressure or high pressure generating on the surface of
Earth or Jupiter. It is expected that flows on the surface of the
companion in a binary system do not form any simple regular pattern 
but become complex flows.  However, the above restrictions allow 
no definite conclusions to be drawn.  Calculations with generalized 
curvilinear coordinates fitting the shape of the companion surface 
are desired.

\section{Conclusions and Discussion}
\subsection{Conclusions}
We perfomed numerical simulations for an ideal gas constituting an
accretion disk in a close binary with mass ratio 1, with a specific
heat ratio of $\gamma=1.01, 1.2$ and compared two-dimensional with
three-dimensional calculations.  From the results obtained, we
conclude as follows.

\begin{enumerate}
\item Whether or not the calculation region may include the companion,
no significant difference appears in the results.

\item An initial condition for gas to be placed at $t=0$ of
$\gamma=1.01$ does not have a large influence, while that of
$\gamma=1.2$ has.

\item A high temperature of gas present on the surface of the
companion gives a thick L1 stream, while a low temperature a fine one;
both do not have very much influence on the structure of the accretion
disk.

\item A specific heat ratio of $\gamma=1.01$ gives, for any parameter,
a neat-shaped accretion disk with spiral shocks.  On the other hand,
with $\gamma=1.2$ the accretion disk formed sometimes takes a greatly
distorted shape depending on the values of parameters.

\item The L1 steam does not reduce its velocity on collision with the
accretion disk or form a hot spot, but forms a bow shock, which can be
called an L1 shock, or a hot line rather than a hot spot.

\item The L1 shock and the penetrating L1 stream enhance the
non-axisymmetry in density distribution in the disk, so that angular
momentum transfer by tidal torque from the secondary star becomes more 
effective.

\item Complex vortices may be present on the surface of the
companion.
\end{enumerate}

\subsection{Discussion}
\subsubsection{Bisikalo et al.'s work}
Attention should be paid here to a series of calculations made by
Bisikalo et al.\cite{rf:32}\tocite{rf:36} \  They performed, like
ourselves, numerical simulations of accretion disks by means of the
finite difference method.  Our calculation scheme differs from theirs
in that we use a far larger number of cells, thus achieving a higher
calculation accuracy, although our calculation region is a little
narrower than theirs.  Nevertheless, their results, varying to some
extent though, are basically similar to ours.  In particular, their
results, like ours, show no hot spots.  They further argue that the
luminosity curve of a close binary can be explained even in the
absence of hot spots.

We also differ from them in the way to explain the mechanism involved
in generation of shocks.  We contend, as before, that spiral shocks
are formed due to the tidal force exerted by the companion, and that
the L1 shock occurs on collision between the L1 stream and the disk
flow.  On the other hand, they do not admit the effect of the tidal
force and, further, propose a mechanism of the shock formation as
caused by collision between the disk flow and the envelope.  The fact
that our results are similar to theirs is still worth noticing,
although the interpretations are somewhat different, and desired to be
checked by a third party.

\subsubsection{Bath et al.'s work}
Bath, Edwards \& Mantle\cite{rf:45} have discussed the penetration 
of the L1 stream into the accretion disk.  According to their 
discussion, whether or not the L1 stream penetrates into the accretion 
disk depends on the magnitude of each of the densities.  That is, an L1
stream having a density larger than that of the accretion disk
penetrates into the accretion disk.  Our calculations just correspond
to this case; the density of our L1 stream is as high as about 10
times that of the accretion disk.  On the other hand, an L1 stream
having a low density will form a hot spot on collision with 
the accretion disk. This
case probably corresponds to the quiescent phase having a hot spot. In
this case the disk temperature is lower than that in the outburst
phase due to the lack of the L1 shock, and the spiral shocks wind more
tightly and they may not be observable by the Doppler tomography
technique.

Bath et al.\cite{rf:45} call the L1 shock, or the hot line we refer to 
in the present paper, which we have found by means of calculations, a
shock-heated wall.  With respect to observation of an actual
penetration, Chochol et al.\cite{rf:46} argue that it is observable in 
the symbiotic star CI Cyg, and Skidmore et al.\cite{rf:47} also argue in
the dwarf nova WZ Sge.

The magnitude of the density of an L1 stream depends on the surface
temperature of the companion.  With a low surface temperature, the L1
stream becomes thin and of low density, so that it may not penetrate
into the accretion disk.

\section*{Acknowledgements}
The authors thank Mr. E. Hayashi for his contribution to preparing a
draft of this paper.  T. Matsuda is supported by Grant-in-Aid for
Scientific Research of Ministry of Education, Science and Culture in
Japan (11134206) and (10640231) of JSPS.  Our calculations were mainly
carried out on NEC SX-4 at the Data Processing Center of Kobe
University and partly on Fujitsu VPP300/16R, VX/4R at the Astronomical
Data Analysis Center of the National Astronomical Observatory, Japan.

\end{document}